# Thermal Contributions to Primordial Nucleosynthesis

Samina Masood and Jaskeerat Singh
Department of Physical and Applied Sciences
University of Houston Clear Lake

## Abstract

Electron mass is known to modify at finite temperatures and densities. Weak nuclear processes have a great impact on electron mass which modifies in a statistical background. We demonstrate how the temperature change in electron mass is associated with beta decay in the early universe. Its precise contributions to the abundance of light elements in the early universe describe some of the details about nucleosynthesis. We employ the calculational scheme of the renormalization of QED to precisely compute the temperature dependence of electron mass during the nuclear processes. In this paper we precisely compute the concentration of electron and its mass change with temperature during nucleosynthesis and use it to describe the helium abundance, expansion rate and energy density of the universe during nucleosynthesis.

## 1. Introduction

Beta decay is a spontaneous process of neutron and contributes to the creation of protons by the release of electrons in the early universe. This decay rate is significantly affected by the electron mass. An overall contribution to beta decay rate is previously computed before and after the nucleosynthesis and is shown to be affected in thermal background. In this paper, we will explicitly show the relation between the electron abundance and helium abundance. The renormalized mass of electron in hot [1-5] and dense [6-10] background of the early universe [11-13] and stars [14-16] behave as affected mass of the theory. In this paper we give a detailed calculation of variation of electron mass and calculate thermal contribution to the concentration of electrons in the early universe and its relation with nucleosynthesis [9-10]. Temperature dependence of electron mass was not the same before and after nucleosynthesis and appears as a complicated function of temperature because of the change in concentration.

We revisit the previously done calculations of thermal corrections of electron composition and temperature dependence of its mass which is evaluated at various temperatures of interest. This generalization will let us figure out how the creation and absorption of leptons can help to understand the process of nucleosynthesis in the early universe. However, thermal contribution to the electron's concentration indicates the end of nucleosynthesis. Temperature dependence of nucleosynthesis parameters such as the rate of production and absorption of electron, helium yield Y and the energy density of the universe are calculated [9-12] as functions of temperature during nucleosynthesis, which explains the dynamics of electrons and its interaction with other particles during the cooling of the universe in the beginning. The change in electron mass in statistical background [9-12] plays a crucial role to determine the detailed mechanism of contributing processes to nucleosynthesis. The statistical background contribution to electron self-mass is calculated up to the two-loop level [4] in literature and determine the range of validity of QED corrections [5], which in turn contributes to helium synthesis.

The radiative corrections to electroweak processes [6] are not significant around nucleosynthesis due to the absence of required concentrations of hot electrons at those temperatures. The statistical contribution to the self-mass of electron is calculated using the renormalization scheme of QED (quantum electrodynamics) in real-time formalism [1]. In this scheme of calculations, the order-by-order cancellation of singularities can be shown by simply adding all of the same order diagrams. Real-time formalism is the only formalism to incorporate thermal effects and still show the cancellation of singularities clearly. We can prove Kinoshita-Lee-Nauenberg (KLN) theorem [17-18] in QED, only in the real-time formalism. KLN theorem emphasizes a need to show order-by-order cancellation of singularities and we could do that in real-time formalism up to the two-loop level. Another benefit of this formalism is that we can separate out thermal corrections from T independent corrections at the first loop level and the higher order contributions can easily be ignored in most cases. At high temperatures, electroweak theory has to be incorporated along with the higher order corrections, even for beta decay. These contributions may be relevant for very early universe, before nucleosynthesis, when temperature is above the mass of the electron (which is of the order of $T>>10^{10}$ K). In the beginning of the universe, electroweak theory and even quantum chromodynamics (QCD) are relevant and even the standard model of high energy physics becomes non-ignorable.

Nucleosynthesis in the early universe took place around T ~1 MeV. The density of the universe [12-14] was low enough ($10^{-31}$g/cc) to be ignored at that time. The mass density of the universe is estimated to be roughly around 9 x $10^{-27}$ kilograms per cubic meter. However, thermal corrections were needed to be incorporated for net concentration of the result of its creation and absorption. Electron concentration is used as indicator to estimate helium synthesis in the early universe. Self-mass of electron clearly has the major thermal correction as neutrino does not interact directly with radiation. Therefore, a detailed study of self-mass of electron during nucleosynthesis is required to estimate the net effect of beta decay and electron capture and its impact on helium synthesis, which is also related to the energy density and the expansion rate of the universe during that period. For a better understanding of self-mass effects, we need to correctly estimate the electron concentration and its self-mass during nucleosynthesis in the early universe.

Next section is devoted to a quick review of the calculation of the exact concentration and self-mass of electron at finite temperature. A detailed study of the electron self-mass during nucleosynthesis shows that the fermion background does not contribute until the temperature reaches close to electron mass. The behavior of electron self-mass navigates the behavior of other parameters in the early universe during nucleosynthesis. We study thermal dependence of these nucleosynthesis parameters in Section 3. The last Section 4 is devoted to the discussion of results in this paper.

## 2. Concentration of Electrons in the Early Universe

The number density of electrons N is calculated from thermal distribution of electrons n(p) with 4-momentum p as:

$$N = \int \frac{d^3p}{(2\pi)^3}\, n(p)$$

Which comes out to be equal to [19]

$$N = \frac{1}{2\pi^2}\left[\frac{m^2}{2\beta}a(m\beta) + \frac{2m}{\beta^2}c(m\beta) + \frac{1}{\beta^3}d(m\beta) - \beta\frac{m^4}{8}g(m\beta)\right] \quad (1)$$

The simple form of Masood's statistical functions then depends on the same exponential function for all a, b,…..g functions in Eq. (2), given as

$$a(m\beta,) = \ln\left(1 + e^{-m\beta}\right), \qquad (2a)$$

$$b(m\beta) = \sum_{n=1}^{\infty} (-1)^n \, Ei(-nm\beta), \qquad (2b)$$

$$c(m\beta) = \sum_{n=1}^{\infty} (-1)^n \frac{e^{-nm\beta}}{n^2} \,. \qquad (2c)$$

$$d(m\beta) = \sum_{n=1}^{\infty} \frac{(-1)^n}{n^3} e^{-nm\beta} \qquad (2d)$$

$$g(m\beta) = \sum_{n=1}^{\infty} (-1)^n [\beta n Ei(-nm\beta) + \frac{e^{-nm\beta}}{m}] \qquad (2e)$$

Using the renormalization scheme of QED, the renormalized mass of electron in a medium is expressed as physical mass, which is given as:

$$m_{phys} = m\,(1 + \frac{\delta m}{m}) \qquad (3)$$

Thermal contributions modify the electron mass such that we can define a physical mass of electron as:

$$\mathrm{m_{phys}} = \mathrm{m} + \delta \mathrm{m}$$

where m is the rest mass of electron and $\delta$m, the self-mass of the electron is the additive contribution due to interaction of electron with radiation in an interacting system. $\delta$m is nonzero even in vacuum [2] computed using the renormalization scheme of statistical QED in real-time formalism. This calculation gives the first order contribution of electron selfmass for a general value of T in the presence of hot fermions. We calculated the dependence of selfmass on temperature and chemical potential $\mu$ in terms of Masood's functions which were first introduced in Ref. [3, 19]. Calculation of the most dominant contributions of the medium (at high energies) to the electron mass for high temperatures and ignorable densities, up to the first order in alpha, has been computed as [3]

$$m_{\rm phys}^2 = m^2 \left[1 - \frac{6\,\alpha}{\pi}\, b\,(m\beta)\right] + \frac{4\,\alpha}{\pi}\, m\, T\, a\,(m\beta) + \frac{2}{3}\,\alpha\,\pi\, T^2 \left[1 - \frac{6}{\pi^2}\, c\,(\,m\beta)\right], \qquad (4)$$

whereas this thermal contribution to $\delta m/m$ up to the first order loop corrections is calculated using the real-time formalism, giving

$$\frac{\delta m}{m}(T) = \frac{\alpha\pi T^2}{3m^2}\,[\{1 - \mathrm{c(m)}\} + \mathrm{b(m\beta)} + \mathrm{a(m\beta)}], \qquad (4)$$

It is worth-noticing that the first term $\frac{\alpha\pi T^2}{3m^2}$ on the right-hand side is due to the interaction of electrons with the radiation whereas Masood's functions (a, b, c…...) appear due to the interaction with hot electrons background. Equations (2) show that the fermion background contribution which is significant after the temperature raises equal to the electron mass or higher. It is indicated that the exponential in Masood's function depends on a dimensionless parameter ($m\beta = m/T$), which increases or reduces exponentially, especially for increasing values of n, as shown in the Table 1. This parameter is due to a comparison between temperature and mass both. Therefore, thermal contributions are significant depending on m$\beta$ for fermions and relevant temperature ranges will be determined by the ratio of temperature with the mass of the fermions [20]. We evaluate $e^{-nm\beta}$ for various values of n showing that the exponential function is steep enough to become ignorable easily. Therefore, the series contributions in Masood's functions are strictly relevant for T~ m and can easily be ignored for smaller (T<< m) and larger (T >>m) values of temperatures easily, even including the value of n. However, the contribution of all of the above functions (Masood's functions) add up together to demonstrate the validity of these functions during nucleosynthesis in the early universe. The general contribution of self-mass of electron in units of the rest-mass of electron is given in Eq. (3). Table 1gives a comparison of values for T << m, and T >> m, with just the contribution of various values of exponential $e^{-nm\beta}$, which allows to truncate the series for the calculation of these functions.

| Table 1: A comparative study of $(T\backslash m)^2$ contribution to T<<m and T>>m which is distinctly different from the simple exponential values | | | | | | | |
|---|---|---|---|---|---|---|---|
| mβ | (T/m) | T << m | T >> m | $e^{-m\beta}$ | $e^{-2m\beta}$ | $e^{-5m\beta}$ | $e^{-10m\beta}$ |
| 10 | 0.1 | 7.65E-05 | 1.15E-04 | 4.54E-05 | 2.06E-09 | 1.93E-22 | 3.72E-44 |
| 9 | 0.111 | 9.44E-05 | 1.42E-04 | 0.000123 | 1.52E-08 | 2.86E-20 | 8.19E-40 |
| 8 | 0.125 | 1.19E-04 | 1.79E-04 | 0.000335 | 1.13E-07 | 4.25E-18 | 1.8E-35 |
| 7 | 0.143 | 1.56E-04 | 2.34E-04 | 0.000912 | 8.32E-07 | 6.31E-16 | 3.98E-31 |
| 6 | 0.167 | 2.12E-04 | 3.19E-04 | 0.002479 | 6.14E-06 | 9.36E-14 | 8.76E-27 |
| 5 | 0.2 | 3.06E-04 | 4.59E-04 | 0.006738 | 4.54E-05 | 1.39E-11 | 1.93E-22 |
| 4 | 0.25 | 4.78E-04 | 7.17E-04 | 0.018316 | 0.000335 | 2.06E-09 | 4.25E-18 |
| 3 | 0.333 | 8.50E-04 | 1.27E-03 | 0.049787 | 0.002479 | 3.06E-07 | 9.36E-14 |
| 2 | 0.5 | 1.91E-03 | 2.87E-03 | 0.135335 | 0.018316 | 4.54E-05 | 2.06E-09 |
| 1 | 1 | 7.65E-03 | 1.15E-02 | 0.367879 | 0.135335 | 0.006738 | 4.54E-05 |
| 0.5 | 2 | 3.06E-02 | 4.59E-02 | 0.606531 | 0.367879 | 0.082085 | 0.006738 |
| 0.1 | 10 | 7.65E-01 | 1.15E+00 | 0.904837 | 0.818731 | 0.606531 | 0.367879 |
| 0.05 | 20 | 3.06E+00 | 4. 59E+00 | 0.951229 | 0.904837 | 0.778801 | 0.606531 |
| 0.01 | 100 | 7.65E+01 | 1.15E+02 | 0.99005 | 0.980199 | 0.951229 | 0.904837 |

The evaluation of self-mass of electron in different regions of temperature helps to understand the behavior of electrons in various statistical environments, which helps to compute the effect of electron mass under statistical conditions at different times in the early universe. The above equations reveal that the self-mass contribution in the extremely hot conditions of the early universe above and below the nucleosynthesis temperatures, with and without including fermion contributions [3, 19], respectively. We can then write the self-mass of electron after nucleosynthesis (T ≤ m), due to the significant radiation background and is derived from the first term in Eq. (2) and is given as:

$$\frac{\delta m}{m} = \frac{\alpha\pi T^2}{3m^2}. \qquad (4a)$$

$$N \sim 0 \qquad (4b)$$

Masood's functions do not contribute to self-mass of electron for T ≤ m. If T is sufficiently smaller than m, then the exponential functions of $m\beta = m/T$ vanishes exponentially. All of the Masood's functions vanish for large values of temperature but equation (2c) can be summed for large T to $\left(c(m\beta)\right) \sim -\frac{\pi^2}{12}$). Therefore, for sufficiently higher temperatures before nucleosynthesis, we use a large T limit such that (T >> m), gives

$$\frac{\delta m}{m} = \frac{\alpha\pi T^2}{2m^2}. \qquad (5a)$$

Whereas the electron concentration can be computed as:

$$N = -\frac{mT^2}{12}\left(1+\frac{\alpha\pi T^2}{2m^2}.\right) \quad (5b)$$

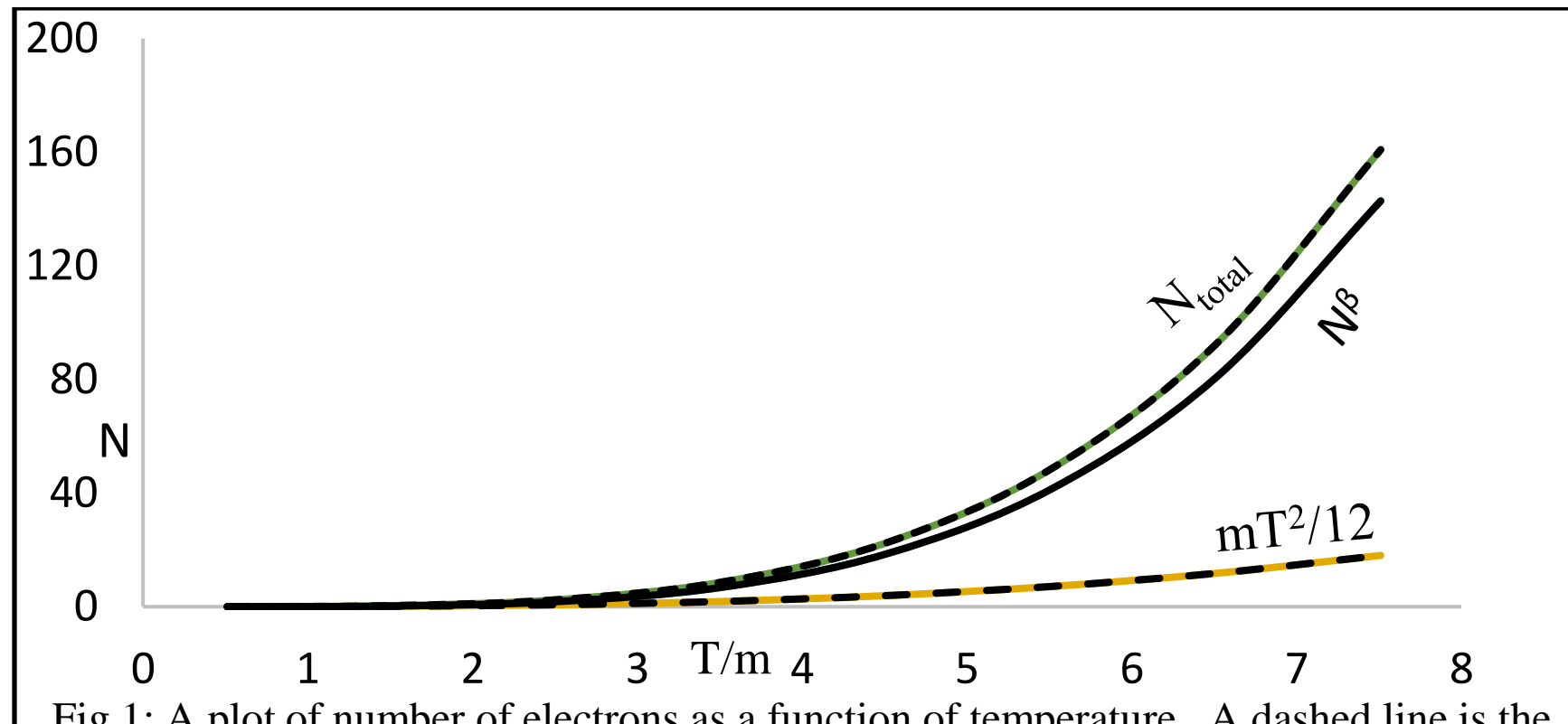

Fig.1: A plot of number of electrons as a function of temperature. A dashed line is the regular value of electron concentration, a solid line is the thermal selfmass correction to the electron concentration and broken line gives the total concentration of electron in the very early universe.

And at extremely high temperatures, it goes to the order of $\frac{T^3}{24\,m}$. Eq. (5b) confirms the presence of free electrons in the early universe as a function of temperature T, and is not necessarily insignificant.

We use this general equation to compute thermal contributions for the entire range of temperatures and plot it as a solid line in Fig. (1). A comparison with the low temperature effect (in the absence of hot fermions background) is calculated from Eq. (4a) and the high temperature contribution (with enough concentration of fermion background) to electron mass, using Eq. (4b). We plot T/m dependence around T ~ m, in general, in Fig.2. However, the range of T in this figure is relevant for the calculation of radiatively created selfmass of any particle of mass. T/m is calculated in Table 2.

Table 2: Thermal contribution to electron self-mass near nucleosynthesis temperatures using "Masood's functions and the approximate values for T < m and T > m in units of m.

| Temp T | Low T Calc) | High T (Calc) | δm /m (consolidated) | δm /m (exact) |
|---|---|---|---|---|
| 0.1 | 0.000292847 | 0.000439271 | 0.000292847 | -0.509896477 |
| 0.2 | 0.001171389 | 0.001757083 | 0.001171389 | -0.13290175 |
| 0.3 | 0.002635624 | 0.003953437 | 0.002635624 | -0.059343751 |
| 0.4 | 0.004685555 | 0.007028332 | 0.004685555 | -0.030333128 |
| 0.5 | 0.007321179 | 0.010981769 | 0.007321179 | -0.01377433 |
| 0.6 | 0.010542498 | 0.015813747 | 0.010542498 | -0.001683382 |
| 0.7 | 0.014349511 | 0.021524266 | 0.014349511 | 0.008703485 |
| 0.8 | 0.018742218 | 0.028113327 | 0.018742218 | 0.018555127 |
| 0.9 | 0.02372062 | 0.03558093 | 0.02372062 | 0.0284381 |
| 1 | 0.029284716 | 0.043927074 | 0.029284716 | 0.038655771 |
| 1.1 | 0.035434507 | 0.05315176 | 0.049383453 | 0.049383453 |
| 1.2 | 0.042169991 | 0.063254987 | 0.06072877 | 0.06072877 |
| 1.3 | 0.04949117 | 0.074236755 | 0.072761123 | 0.072761123 |
| 1.4 | 0.057398044 | 0.086097065 | 0.085527131 | 0.085527131 |

| 1.5 | 0.065890611 | 0.098835917 | 0.099059208 | 0.099059208 |
|---|---|---|---|---|
| 1.6 | 0.074968873 | 0.11245331 | 0.113380567 | 0.113380567 |
| 1.7 | 0.08463283 | 0.126949244 | 0.128508261 | 0.128508261 |
| 1.8 | 0.09488248 | 0.14232372 | 0.144455104 | 0.144455104 |
| 1.9 | 0.105717825 | 0.158576738 | 0.161230912 | 0.161230912 |
| 2 | 0.117138865 | 0.175708297 | 0.178843338 | 0.178843338 |

It can be seen that the exact calculation from Eq. (1) is needed below T< 2MeV. Selfmass corrections are proportional to $\alpha$, which is another regulating parameter. It still causes the selfmass corrections to move higher than 1 which makes the perturbation theory invalid. However, Eq. (2) gives unphysical results at T < m because of the absence of free electrons in the medium. The Masood's functions seem to be required to consolidate the low temperature and high temperature results. Eq. (1) reproduces the results of T >> m calculations above T > 2.5 MeV. Whereas the results differ significantly around T/m =1 (T = m = 0.511MeV) and low temperature results are also reproduced to a good approximation. Whereas, T ~ m is relevant for nucleosynthesis in the early universe as beta decay and other nuclear processes involving electrons and positrons contributed significantly to the primordial nucleosynthesis. However, thermal corrections remain the same for electrons and positrons in the early universe. Electrons in the beginning of the universe were significantly contributed by pair production or through other leptonic processes. Whereas the weak processes contribute below the neutrino decoupling temperatures, especially in the early universe where we can easily ignore the density effects.

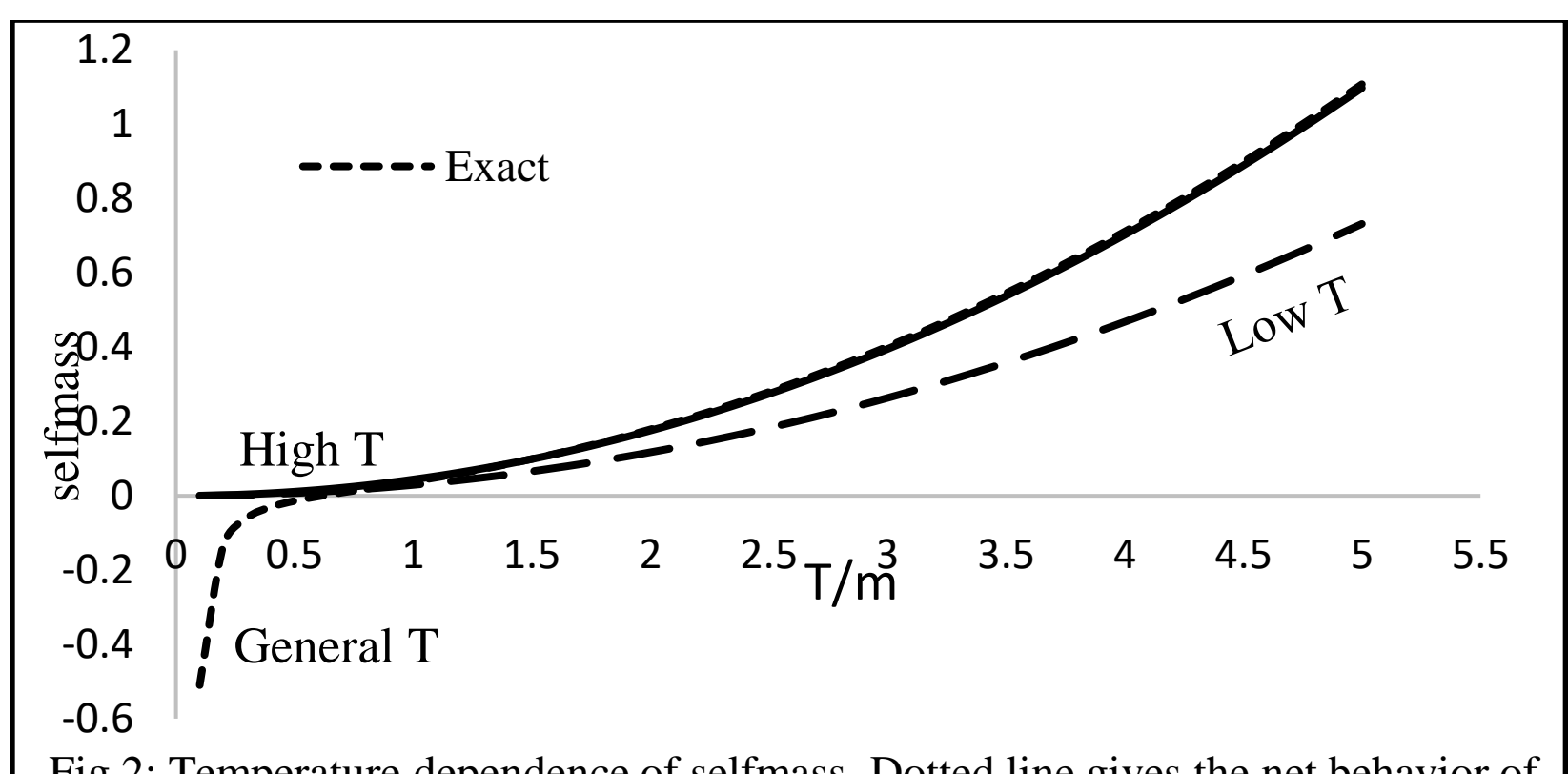

Fig.2: Temperature dependence of selfmass. Dotted line gives the net behavior of mass with T. Dashed line correspond to low temperature T < m behavior with the particle mass, whereas the solid line corresponds to selfmass at T>m

Fig. (2) shows a direct comparison of high temperature and low temperature behavior of the electron self-mass with the generally computed self-mass values from Masood's functions. As mentioned earlier, the negative contribution of Masood's functions reduced self-mass for sufficiently small T << m, thus revealing that free hot fermions indicate the absence of fermion interaction. Fig.(2a) clearly shows the comparative selfmass behavior near nucleosynthesis. Dotted line in this figure gives the exact value of selfmass. Exponential dependence of the fermion distribution function contributes negatively to overcome thermal correction of photon background ($\frac{\alpha\pi T^2}{3m^2}$), giving an overall negative result until T > m. For lower temperatures the electrons do not stay free and total mass of electron may seem to be reduced for bound electrons. This behavior is changed for T slightly greater than m. The significance of Masood's functions is to describe the thermal behavior during nucleosynthesis, mainly.

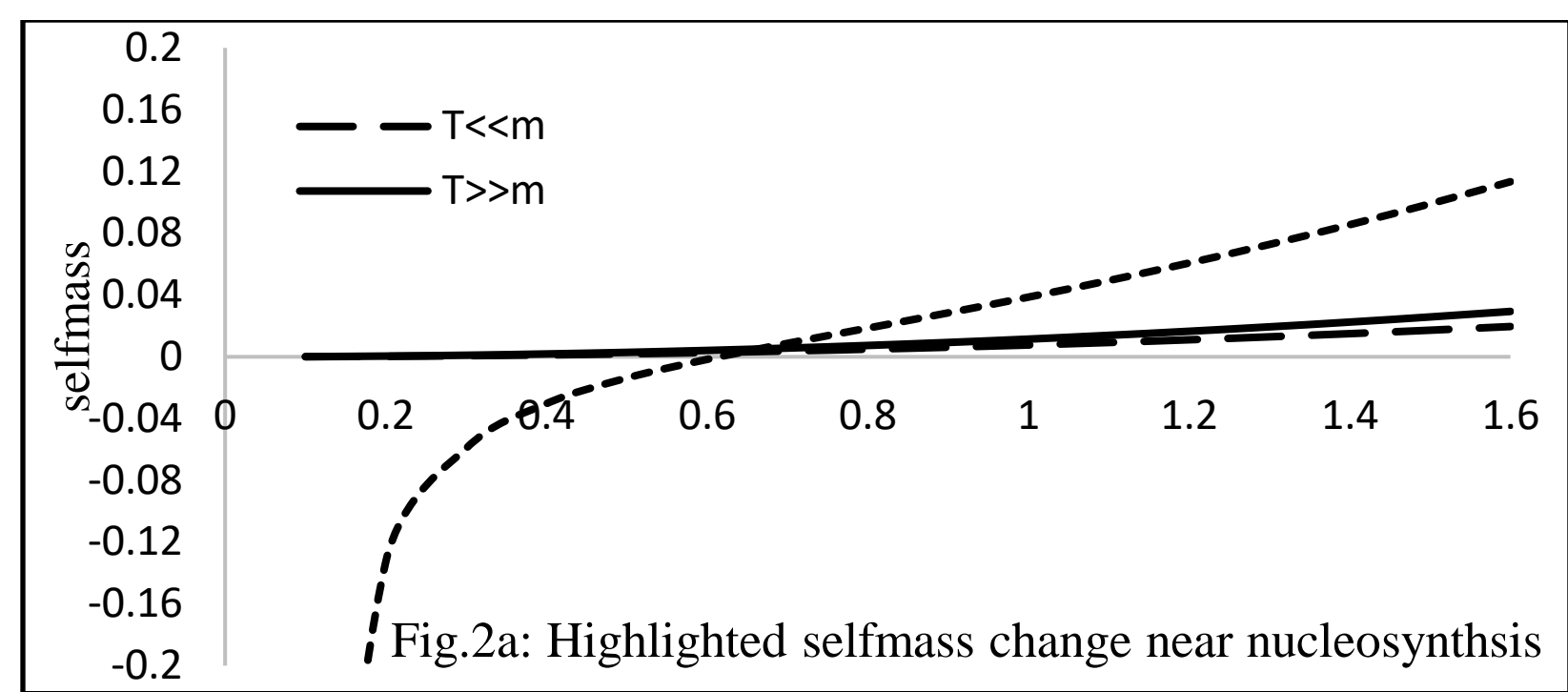

Fig.2a: Highlighted selfmass change near nucleosynthsis

Self-mass below the temperature of electron mass is almost a straight line and is only a very small fraction of the particle's rest mass, whereas the selfmass above the electron mass increases exponentially as Masood's functions for some time until the temperature is too high for nucleosynthesis. Above T = 2 MeV (the neutrino decoupling temperature), the nucleosynthesis stops and selfmass increases quadratically on the ratio between temperature and electron mass ($\sim T^2 / m^2$). When the nucleosynthesis starts at $T \leq 2$ MeV, the rise in electron temperature is decreased as it is associated with the burning of primordial hydrogen into helium. This graph shows a significant decrease in thermal corrections after the nucleosynthesis stops. Table 2 shows a comparison of calculated values of self-mass for various temperatures at T/m greater than 1, indicating that the dominant high temperature contribution is a quadratic function of T as given in Eq. (4b). These corrections are extremely small for $T < 0.1m$ and reaches up to unity very quickly after $T/m > 2$ and it increases almost exponentially during nucleosynthesis.

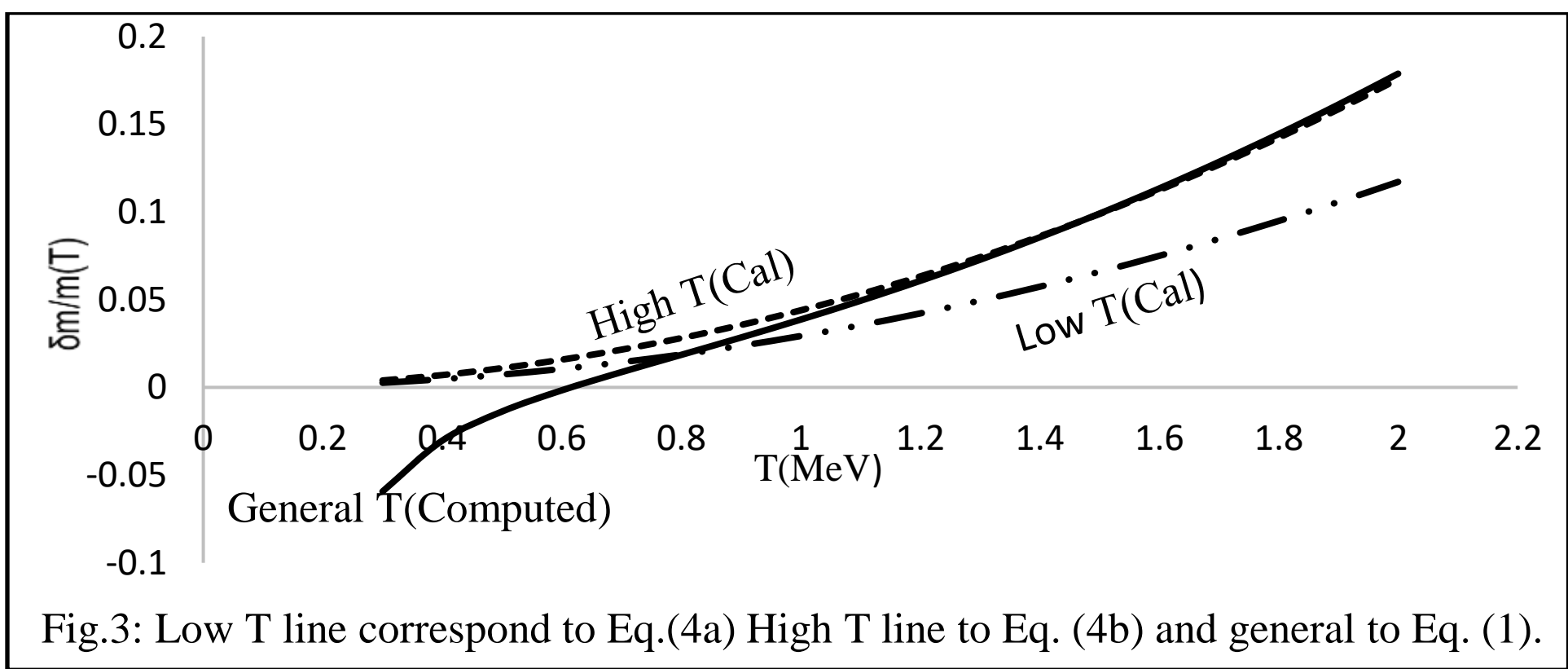

Fig.3: Low T line correspond to Eq.(4a) High T line to Eq. (4b) and general to Eq. (1).

A plot of these functions in Fig. (3) highlights the fermion contribution from the medium during nucleosynthesis and shows clearly where two graphs coincides (overlaps) at high T which is around T=1.2 MeV. This is around 2.5 times the electron mass. At this point, fermion contribution is almost comparable to photon background [3] and contribution of $c(m\beta)$ dominates over the contributions of $b(m\beta)$ and $a(m\beta)$. However, Masood's functions describe the functional dependence of self-mass and the relevant impact on the beta decay rates during nucleosynthesis. In this way, electron mass, beta decay rate, and helium abundance change with respect to their vacuum values, and all varies with temperature. A consolidated graph of self-mass as a function of temperature is shown in Fig. (4), where we calculate the low temperature contribution from Eq. (4a) and high temperature contribution from Eq. (4b). A disconnected region below 0.6 MeV shows the point where low temperature and high temperature behavior starts to differ. This is the region where nucleosynthesis actually took place. For a heating system, going from lower temperature, neutrino emission starts due to beta decay processes. This is the region that is studied in detail in this paper using Masood's functions.

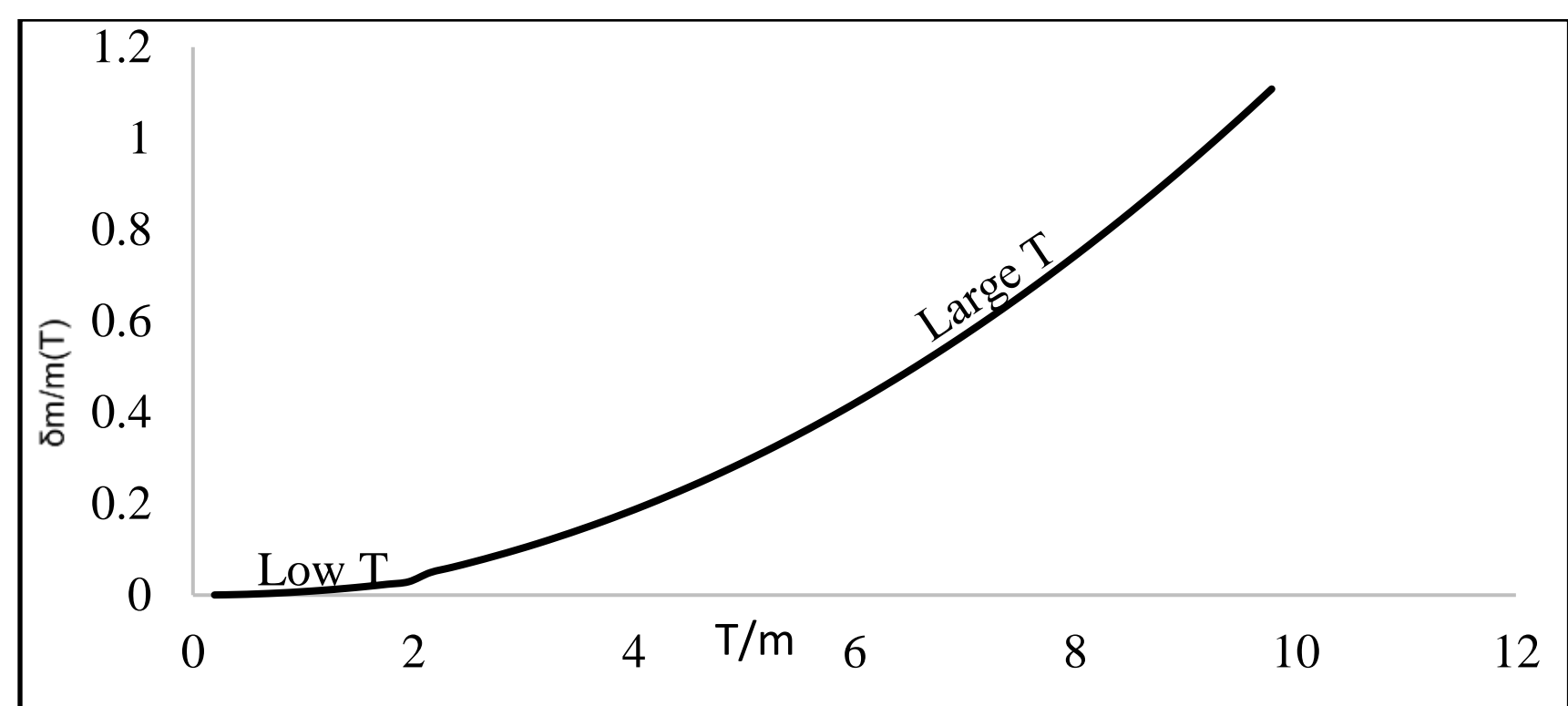

Fig. 4: Thermal correction to self-mass of lepton as a function of temperature using Eqs. (4a and 4b) in the relevant ranges. m corresponds to lepton mass and T will vary for every value of m. It is a consolidated graph for T dependent part of mass.

The computed contributions use Masood's functions, or in other words, include the fermion contribution through Masood's functions. Whereas the low temperature contribution is mainly coming from the interaction of radiation with electrons only. Degenerate pressure of fermions may not allow the creation of electrons at low temperature beyond certain limit. That explains why the selfmass reduces the physical mass at temperatures below the electron mass. Therefore, at low temperature, the presence of electrons reduces the selfmass instead of increasing. It means that at low temperatures, the free electrons are not allowed to stay independently.

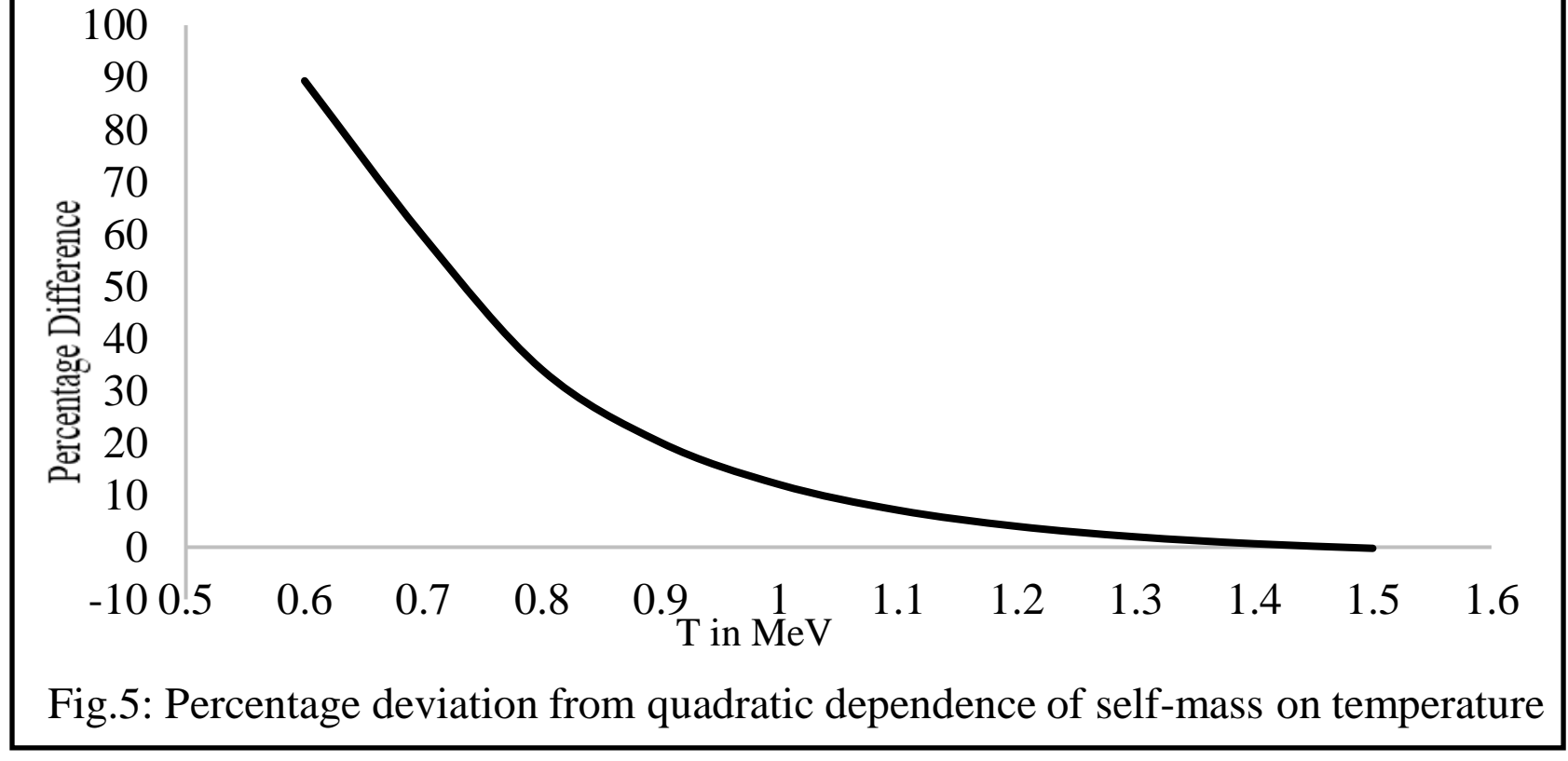

Fig.5: Percentage deviation from quadratic dependence of self-mass on temperature

We have compared the contributions of Masood's functions during nucleosynthesis, as compared to the high temperature expansion of Eq. (4b) for T >> m, prior to the start of nucleosynthesis in the early universe. It shows that the significance of self-mass corrections depends a lot on Masood's functions during nucleosynthesis and their contribution is required for the precise calculations of the decay rate. All these values are plotted in Fig. (5) which shows that the self-mass contribution via Masood's functions is maximized around the temperatures near the electron mass and just approximately reduced to quadratic dependence of T/m around 1.5 MeV. Fig. (6) indicates the temperature (below neutrino decoupling temperature $T < 2$ MeV) where nucleosynthesis started to contribute, significantly, to helium synthesis in the early universe. The overall behavior of the self-mass of electrons, as a function of temperature for 0.6 MeV $< T <$ 1.6 MeV, is plotted in Fig. (6) to closely study nucleosynthesis. Fermion background starts to add thermal mass around 0.6 MeV. The increase in mass decreases the decay rate according to the regular calculations of decay rates.

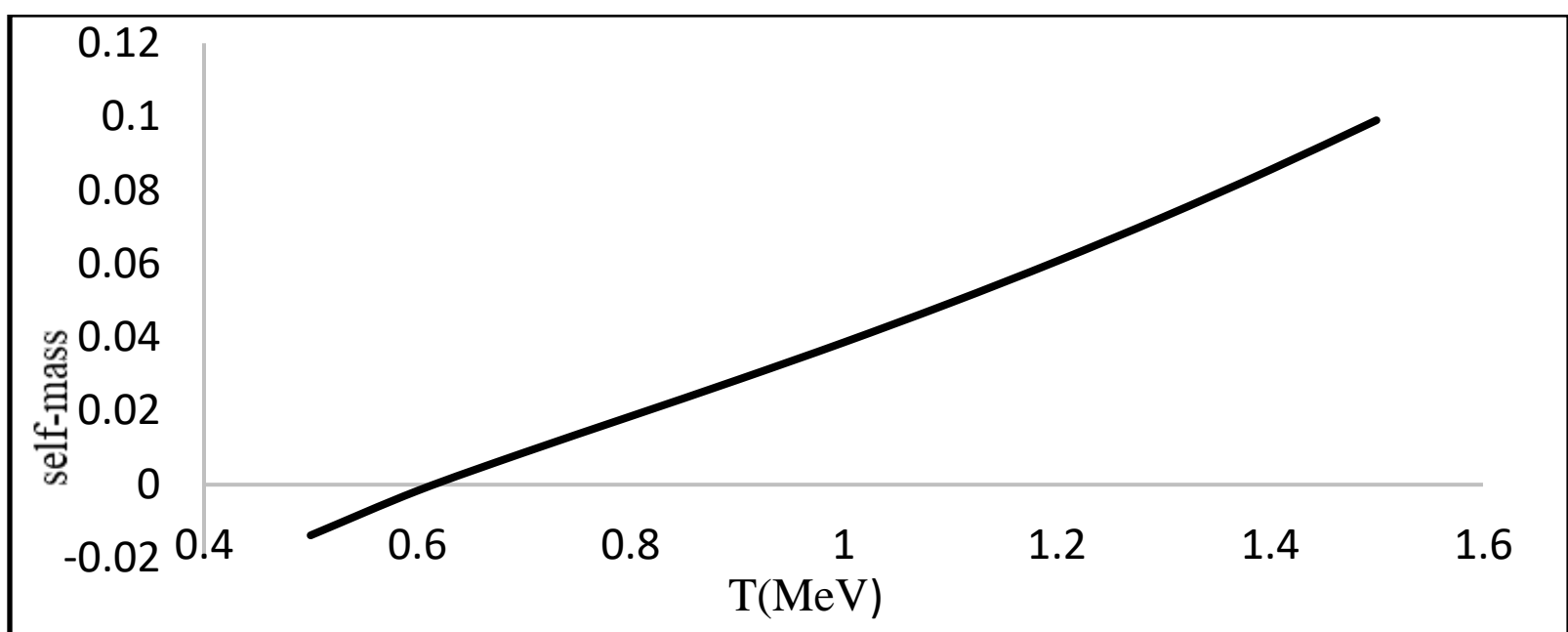

Fig. 6: Plot of the self-mass of electron during nucleosynthesis. Increase of self-mass with temperature corresponds to decrease in physical mass of electron with cooling of the universe.

In the next section, we study the impact of temperature on nucleosynthesis parameters including beta decay and helium yield in the early universe. We can also see the impact of mass-change on the energy density of the universe and the Hubble parameter. Temperature dependence is induced to all related parameters through the thermal mass of electron. Beta decay processes are possible for higher leptonic generations, but they are suppressed by the large masses of the muon and tauon [20], even in vacuum. Thermal contributions come from the medium and the existence of enough concentration of fermions depend on their mass. The existence of enough free muons and heavier particles cannot be ruled out in the very early universe where temperatures were possible to express in muon mass and tauon mass. However, nucleosynthesis does not occur there. Similarly, the mass of the proton is taken constant as its direct interaction with radiation and charge is significantly suppressed by heavy baryonic mass near nucleosynthesis.

### 3. Temperature Dependence of Nucleosynthesis Parameters in the Early Universe

It is well-understood that the increase in electron mass reduces the decay rate for heavier mass and increased temperature decreases the possibility of decay rate and hence decreases the rate of nucleosynthesis. The beta decay rate in the early universe is proportional to the thermally corrected self-mass of the electron.

Thermal contributions to electron mass are found to be related to beta decay rate $\frac{\delta\lambda}{\lambda}$, and hence the helium abundance $\frac{\delta Y}{Y}$ in the early universe. There are several processes that can contribute to net beta decay such as neutrino capture or inverse beta decay [11-12].

$$\frac{\delta\lambda}{\lambda} = -0.2\frac{\delta m}{m}\left(\frac{m}{T}\right)^2 \qquad (6)$$

$$\frac{\delta Y}{Y} = -0.2\frac{\delta\lambda}{\lambda} = 0.04\frac{\delta m}{m}\left(\frac{m}{T}\right)^2 \qquad (7)$$

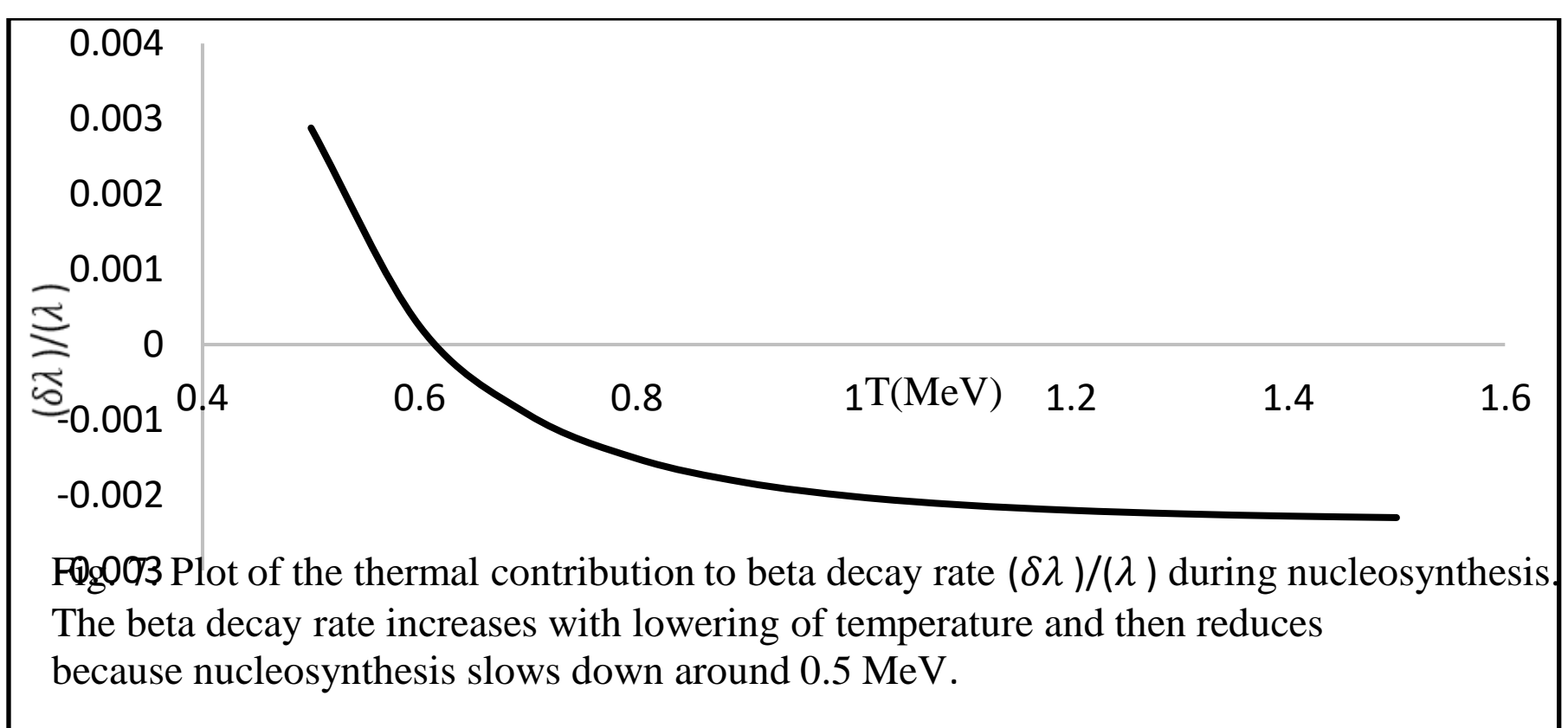

Fig. 7: Plot of the thermal contribution to beta decay rate $(\delta\lambda)/(\lambda)$ during nucleosynthesis. The beta decay rate increases with lowering of temperature and then reduces because nucleosynthesis slows down around 0.5 MeV.

Eqs. (6) and (7) show that the beta decay rate, and the helium abundance parameter are proportional to the self-mass and vary relatively slowly with temperature, respectively. These functions are plotted against temperature in Fig (7). It shows that a plot of thermal contribution to beta decay rate as a function of temperature. However, thermal effects to helium synthesis are not necessarily ignorable. It is known that helium yield is around 25% in the early universe. Therefore, 0.003/0.25 gives a contribution between 1-2 percent after the completion of nucleosynthesis. A plot of thermal corrections to helium yield Y is given in Fig. (8)

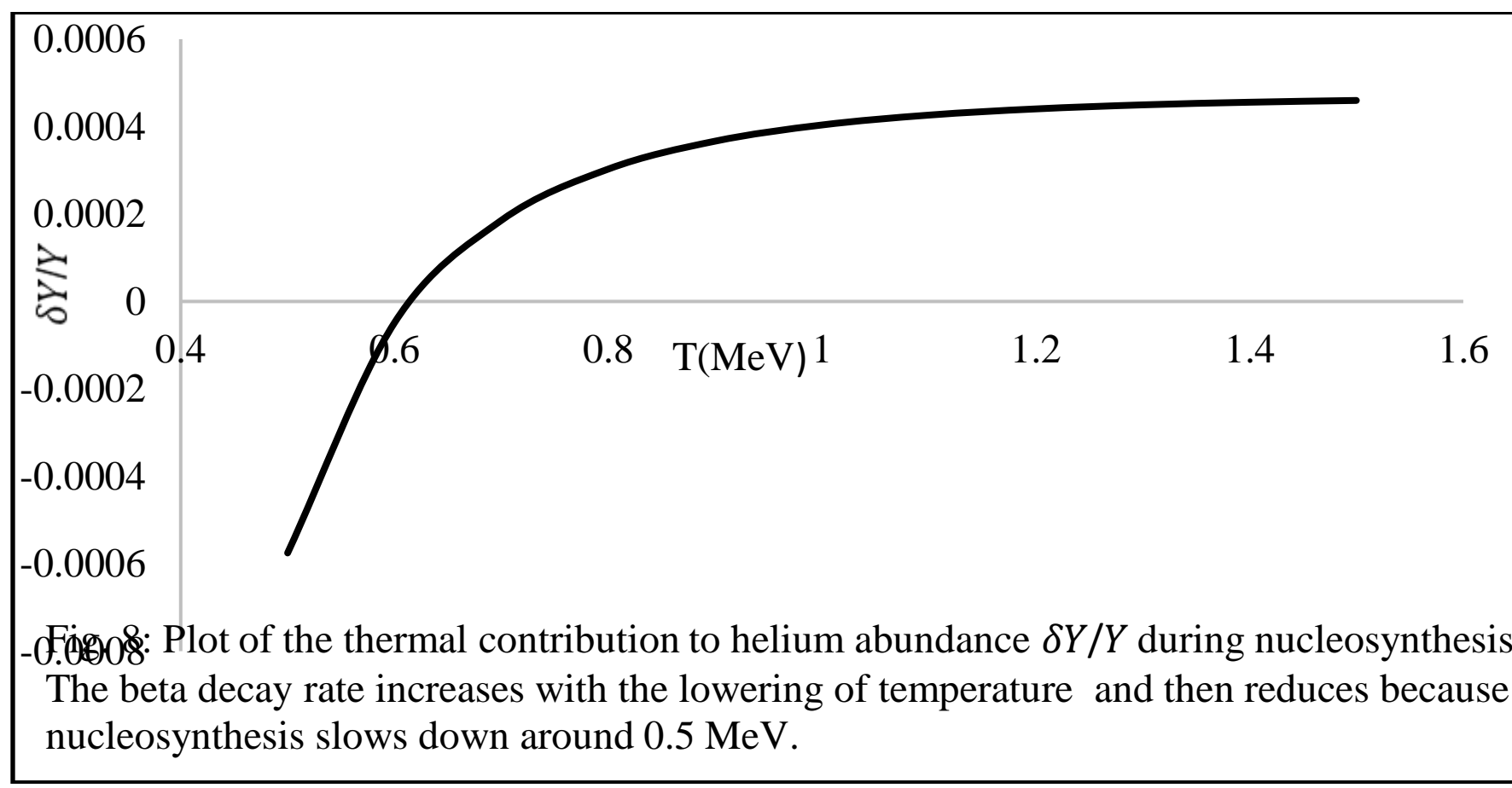

Fig. 8: Plot of the thermal contribution to helium abundance $\delta Y/Y$ during nucleosynthesis The beta decay rate increases with the lowering of temperature and then reduces because nucleosynthesis slows down around 0.5 MeV.

On the other hand, the modification in the energy density of the universe with respect to temperature becomes:

$$\frac{\delta\rho}{\rho} = -\frac{\delta m}{m}\left(\frac{m}{4T}\right)^2 \qquad (8)$$

The plot of the variation in energy density of the universe during nucleosynthesis is shown in Fig. (9). This shows the increase in energy density with the decrease in temperature. However, energy density is independent of temperature before or after nucleosynthesis as well.

The Hubble constant H changes with respect to temperature as compared to the calculated [11] value of the Hubble constant H as:

$$\frac{\delta H}{H} = -0.5\frac{\delta m}{m}\left(\frac{m}{4T}\right)^2 \qquad (9)$$

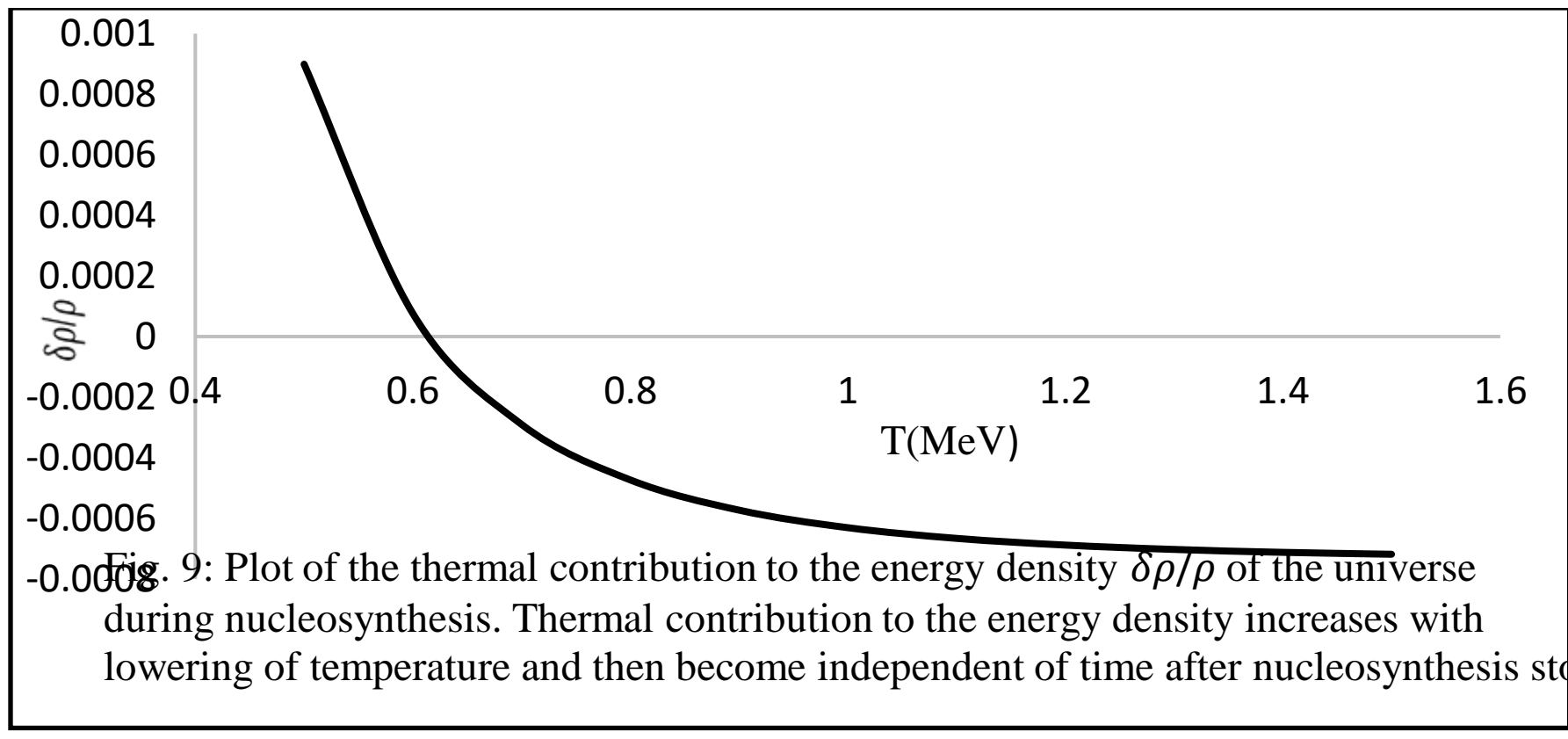

Fig. 9: Plot of the thermal contribution to the energy density $\delta\rho/\rho$ of the universe during nucleosynthesis. Thermal contribution to the energy density increases with lowering of temperature and then become independent of time after nucleosynthesis st

These plots show that the electron background introduces thermal contribution to nucleosynthesis due to the dependence of electron mass on temperature. Their contribution can be calculated comparing the concentration of electrons before and after nucleosynthesis. This effect is determined on the helium yield and then the beta decay rate, energy density of the universe and then the Hubble parameter.
Similar behavior is expected by the Hubble parameter as well and is given by Fig. (10), which gives a plot of Eq. (9).

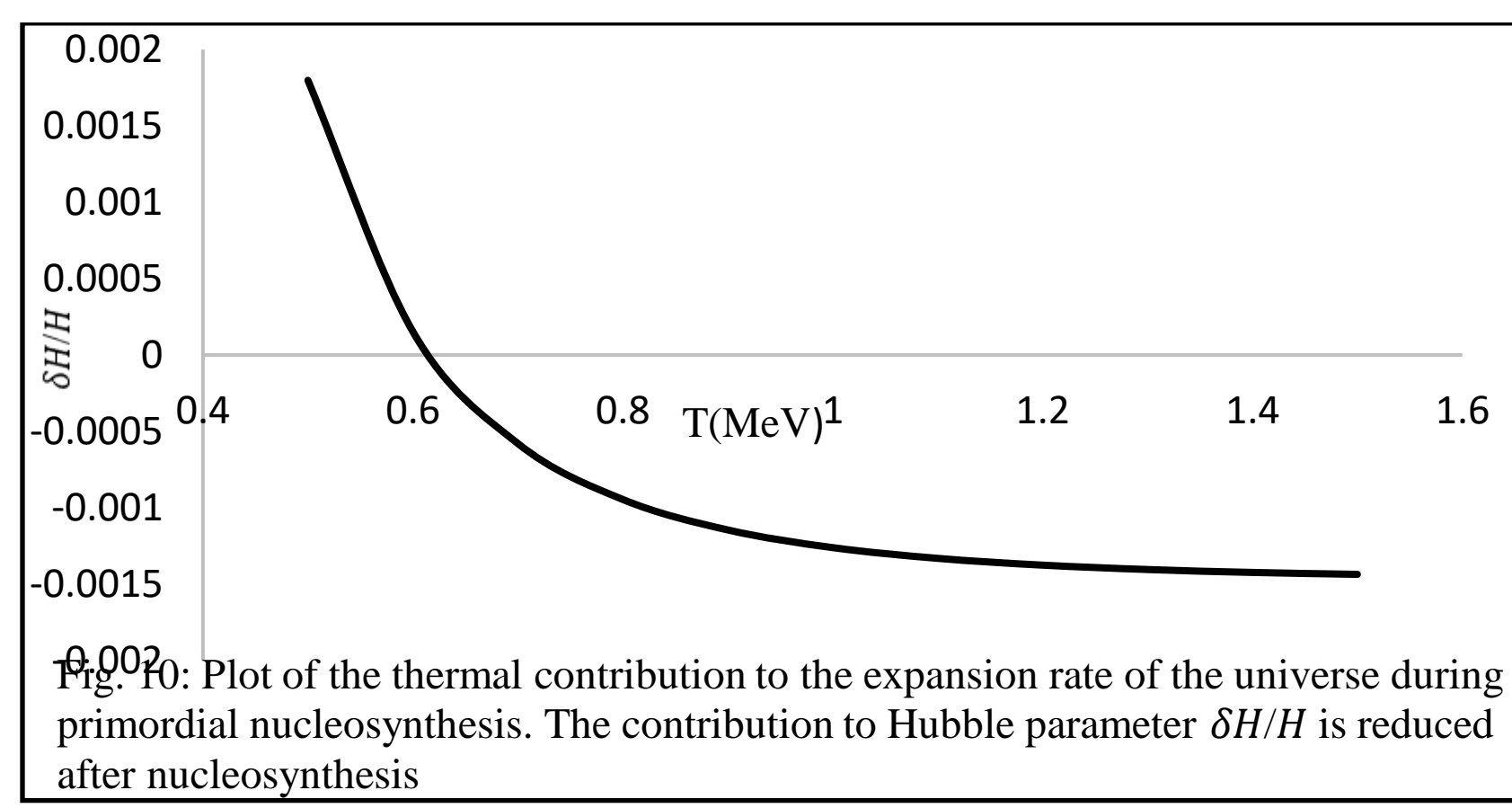

Fig. 10: Plot of the thermal contribution to the expansion rate of the universe during primordial nucleosynthesis. The contribution to Hubble parameter $\delta H/H$ is reduced after nucleosynthesis

## 4. Results and Discussions

A detailed study of the electron self-mass during nucleosynthesis shows that the fermion background cannot be there even around the temperatures close to electron mass. At T≤ m, electrons interact with the hot photons in the background only. The effect of fermion background gives an additive contribution at higher temperatures, that is slightly above the electron rest mass (0.511 MeV), i.e; T > 0.6 MeV

Thermal contributions during nucleosynthesis in the early universe are mainly expressed in terms of Masood's functions. High temperature and low temperature values of the beta decay rate and the helium abundance parameters, before and after nucleosynthesis, become independent of temperature. Eqs. (6-9) reduce to the following equations after cancellation of temperature dependence before nucleosynthesis as:

$$\frac{\delta\lambda}{\lambda} = -0.2\frac{\alpha\pi}{2} \qquad (10a)$$

$$\frac{\delta Y}{Y} = 0.04\frac{\alpha\pi}{2} \qquad (10b)$$

$$\frac{\delta\rho}{\rho} = -\frac{\alpha\pi}{48} \qquad (10c)$$

$$\frac{\delta H}{H} = -\frac{\alpha\pi}{16} \qquad (10d)$$

After the nucleosynthesis is stopped in the early universe, we obtain the following set of equations:

$$\frac{\delta\lambda}{\lambda} = -0.2\frac{\alpha\pi}{3} \qquad (11a)$$

$$\frac{\delta Y}{Y} = 0.04\frac{\alpha\pi}{3} \qquad (11b)$$

$$\frac{\delta\rho}{\rho} = -\frac{\alpha\pi}{32} \qquad (11c)$$

$$\frac{\delta H}{H} = -\frac{\alpha\pi}{12} \qquad (11d)$$

Actual changes in the values of nucleosynthesis parameters and the contribution to the rate of expansion due to the hot fermion background in the early universe are tabulated in Table (3). Eqs. (10-11) show that the nucleosynthesis parameters change before and after nucleosynthesis. However, they do not remain temperature dependent any more as shown by the flatness of all graphs in Figs (7-10) below and above nucleosynthesis. However, the temperature dependence is not totally removed during nucleosynthesis. A few constant values of these parameters are shown in Table (3) for comparison.

Table 3: Values of nucleosynthesis parameters, before and after

| Quantity | Symbol | Before Nucleosynthesis | After Nucleosynthesis |
|---|---|---|---|
| Electron Self-mass | $\frac{\delta m}{m}$ | $\frac{\alpha\pi T^2}{2m^2}$ | $\frac{\alpha\pi T^2}{3m^2}$ |
| Beta Decay Rate | $\frac{\delta\lambda}{\lambda}$ | -0.002294056 | -0.001529371 |
| Helium Abundance | $\frac{\delta Y}{Y}$ | 0.000458811 | 0.000305874 |
| Energy Density | $\frac{\delta\rho}{\rho}$ | -0.0007166893 | -0.000477928 |
| Hubble Parameter | $\frac{\delta H}{H}$ | -0.000358446 | -0.000238964 |

It shows that considering the change in temperature and the composition of the hot fermion background without estimating thermal effects at a particular temperature that is not the same at T < m and T > m. It can then be calculated that all the nucleosynthesis parameters change before and after nucleosynthesis and this change occurs during nucleosynthesis.

Thermal contribution to the self-mass of the electron contribute to nucleosynthesis and then energy density of the universe. However, thermal contribution is very small as expected due to the slowly varying functional dependence of Masood's functions on temperature due to its alternating series.
Our results show that the fermionic concentration changes significantly during nucleosynthesis during nucleosynthesis regardless of slow temperature dependence of beta decay. This is totally understandable that why thermal dependence is small during nucleosynthesis.

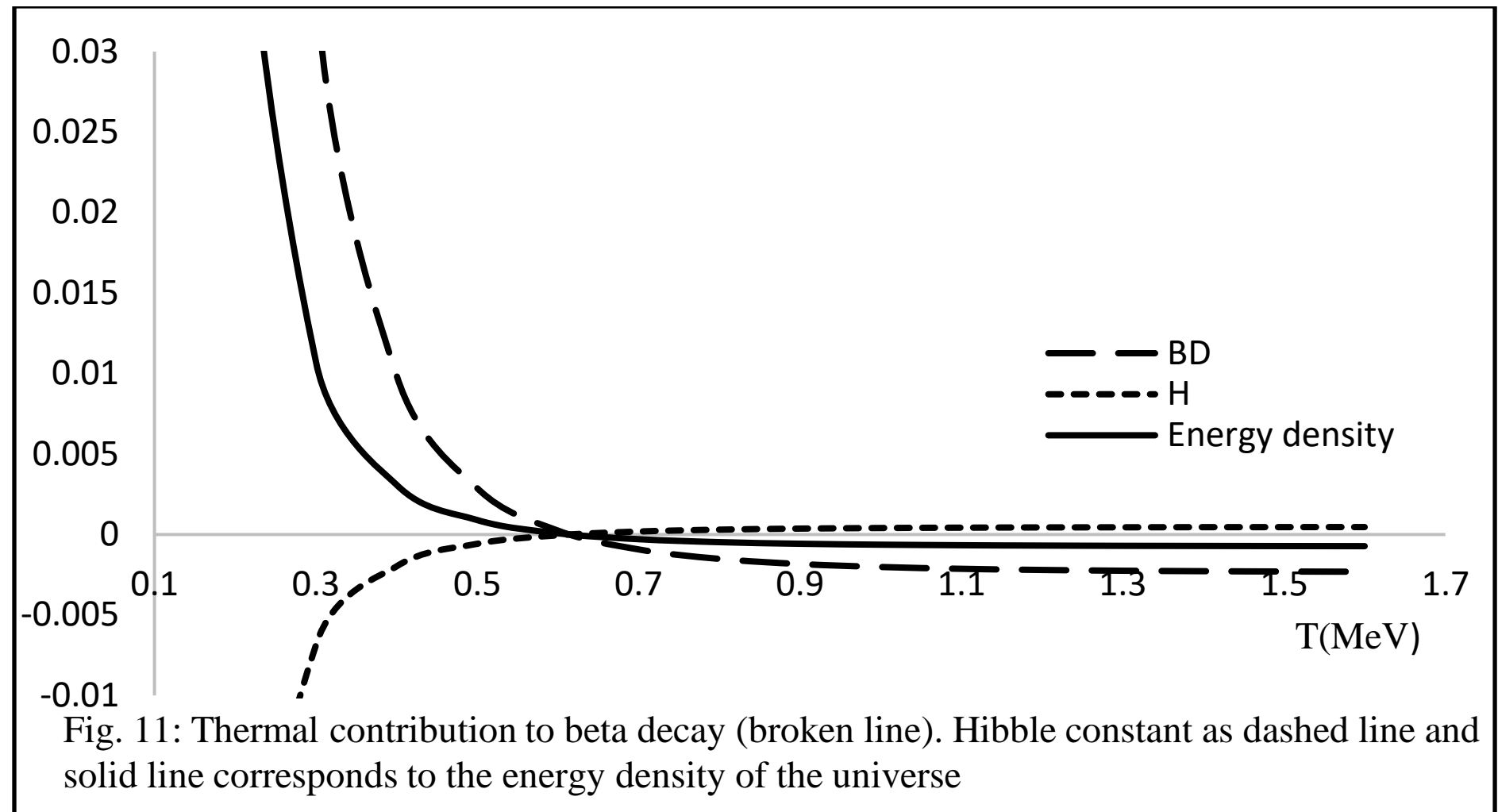

Fig. 11: Thermal contribution to beta decay (broken line). Hibble constant as dashed line and solid line corresponds to the energy density of the universe

Helium abundance is associated with the conversion of proton into neutron by the absorption of electrons which causes the consumption of electrons and is indicated by the reduction of number of electrons during nucleosynthesis associated with thermal dependence of N. The value of N can be used as a useful parameter to understand the composition of the universe. Negative thermal contribution to N may arise due to the reduction of free electrons or presence of bound electrons.

This analysis of calculation of N can help to understand the change in composition of stellar cores at various conditions of temperatures and density. Stellar nucleosynthesis is not a new topic. Its study started back in the early 20$^{th}$ century [21] and can be found in literature regularly [22-26]. However, we can use the approach of this paper for stellar nucleosynthesis as well.